# A New Graphical Password Scheme Resistant to Shoulder-Surfing

Haichang Gao, Zhongjie Ren, Xiuling Chang,
Xiyang Liu
Software Engineering Institute
Xidian University
Xi'an, Shaanxi 710071, P.R.China
hchgao@xidian.edu.cn

Uwe Aickelin
School of Computer Science
The University of Nottingham
Nottingham, NG8 1BB, U.K.
uxa@cs.nott.ac.uk

*Abstract*—Shoulder-surfing is a known risk where an attacker can capture a password by direct observation or by recording the authentication session. Due to the visual interface, this problem has become exacerbated in graphical passwords. There have been some graphical schemes resistant or immune to shoulder-surfing, but they have significant usability drawbacks, usually in the time and effort to log in. In this paper, we propose and evaluate a new shoulder-surfing resistant scheme which has a desirable usability for PDAs. Our inspiration comes from the drawing input method in DAS and the association mnemonics in Story for sequence retrieval. The new scheme requires users to draw a curve across their password images orderly rather than click directly on them. The drawing input trick along with the complementary measures, such as erasing the drawing trace, displaying degraded images, and starting and ending with randomly designated images provide a good resistance to shoulder-surfing. A preliminary user study showed that users were able to enter their passwords accurately and to remember them over time.

## I. INTRODUCTION

Graphical passwords have been proposed as a useful authentication method for Personal Digital Assistants (PDAs) which are increasingly used with their small size, compact deployment and low cost [1]. Given the fact that pictures are generally easier to remember than words [2, 3] and that humans are the 'weakest link' in any password authentication mechanism [4-6], it is conceivable that graphical passwords would be able to provide a good trade-off between usability and security.

However, most of the current graphical password schemes are vulnerable to shoulder-surfing [7-10], a known risk where an attacker can capture a password by direct observation or by recording the authentication session. Due to the visual interface, shoulder-surfing becomes an exacerbated problem in graphical passwords. Several approaches have been developed to deal with this problem, but they have significant usability drawbacks, usually in the time and effort to log in, making them less suitable for everyday authentication [11-13]. For example, it is time-consuming for users to log in CHC [11] and there are complex text memory requirements in scheme proposed by Hong [19]. With respect to the scheme proposed by Weinshall [20], not only is it intricate to log in, but also the main claim of resisting shoulder-surfing is proven false [23]. In this paper, we introduce a new graphical password scheme which provides a good resistance to shoulder-surfing and preserves a desirable usability.

Our inspiration comes from two representative graphical password schemes: DAS [7] and Story [9]. DAS allows users to draw a free-form picture on $N \times N$ grid to produce a password [7] and Story requires users to select a sequence of images to make a story [9]. Our new shoulder-surfing resistant scheme CDS (Come from DAS and Story) adopts a similar drawing input method in DAS and inherits the association mnemonics in Story for sequence retrieval. It requires users to draw a curve across their password images (pass-images) orderly rather than click directly on them. The drawing method seems to be more compatible with people's writing habit, which may shorten the login time. The drawing passes through both pass-images and decoys, which used to confuse peepers. To avoid revealing the first and last pass-images, the drawing must begin and end with given random images. To enhance its shoulder-surfing resistant properties further, CDS displays degraded images which are difficult to distinguish from a distance or from a side view. Moreover, the majority of the drawing trace will be cleared away as the stylus being sliding, reducing the probability of passwords being revealed. Other complementary measures, such as limiting the length of drawing trace, are also deployed to strengthen the security.

We conduct a user study to explore the usability of CDS in terms of accuracy, efficiency and memorability, and benchmark the usability against that of a Story scheme. The result was encouraging that novice users were able to enter their passwords accurately and to remember them over time. It took about 50 seconds on average to create a password and construct a story. In a five-minute test, the mean time to log in was 13.7 seconds and one week later all the participants could recall their CDS passwords correctly and took 19.8 seconds on average to log in. In comparison to Story, our scheme had a similar memorability, probably due to the same association mnemonics. In terms of login time, CDS was not as good as Story and the degraded images in CDS were most likely to impinge on its efficiency. At the same time, there was a remarkable downtrend in the time to log in

CDS as the number of login increased, indicating that the login time will decreases with familiarization.

The remainder of the paper is organized as follows: The following section briefly reviews some exemplars of graphical password schemes. Section 3 gives a detailed introduction to our prototype of CDS. Section 4 presents the preliminary experiment and result. Section 5 discusses additional observations and possible extension to our scheme. Conclusion and future work are addressed in Section 6.

## II. RELATED WORKS

The trend toward a highly mobile workforce and the ubiquity of graphical interfaces (such as the stylus and touch-screen) has enabled the emergence of graphical authentications in PDAs. According to the memory task involved in remembering and entering the password, graphical passwords can be divided into three general categories: recall-based systems, cued-recall systems and recognition-based systems [13, 14].

DAS is the first recall-based graphical password scheme in which the password is a free-form picture drawn on a 2D grid [7]. This scheme releases users from remembering complex text string and has a large theoretical password space. But the drawing rules are difficult to follow, resulting in a usability problem. For example, users cannot locate strokes too close to a grid-line or cross a corner. To solve this issue, Pass-Go (named after an ancient board game Go) is proposed which allows users to draw their password using grid intersection points instead of grid cells in DAS [15]. Another modification to DAS is YAGP [24] where approximately correct drawing can be accepted by dividing "trend quadrants" and adopting Levenshtein distance string matching.

Cued-recall systems typically require users to remember and target specific locations within a presented image. A well-known scheme in this category is PassPoints which has attracted great attention [10]. In this scheme, users should choose several points on an image and click orderly on them within a tolerance for authentication. Security analyses find it vulnerable to hotspots and simple patterns within images [13, 17, 18]. A commercial version of PassPoints for the PocketPC is available from visKey for screen-unlock [16].

In recognition-based systems, users must recognize the target images embedded amongst a set of distractor images. Passfaces [25], a representative scheme of this category, is based on face recognition and has a high degree of memorability [26, 27]. Davis finds that people tended to select faces of their own race and gender and presents a similar scheme Story which replaces faces with classified objects. Story introduces a sequential component to improve security and utilizes an association mnemonics for sequence retrieval. In specific, users must memorize their images in order, aided by constructing a story to connect them.

A common vulnerability to the above systems is shoulder-surfing, a well known method of stealing passwords. It occurs when an attacker learns a user's password by watching the user log in. To overcome this problem, many shoulder-surfing resistant schemes have been devised [11, 19, 20, 28]. In CHC [11], users need to identify their pass-objects, visualize the triangle they form and click inside the convex hull. Unlike CHC[11], the shoulder-surfing resistant graphical scheme proposed by Man, et al [19] depends on users remembering several images (pass-objects) and the their corresponding text codes as well as coding the relative location of the pass-objects in reference to a pair of eyes. In the scheme proposed by Weinshall [20], it involves remembering a number of images and computing a path through a panel of images according to certain rules. ColorLogin uses background color to decrease login time and resist shoulder-surfing [28]. All of them can provide a good protection against shoulder-surfing. However, a user study reveals that their inherent drawbacks in usability make them less suitable for everyday authentication [13]. Motivated by this, we design a new shoulder-surfing resistant scheme CDS which is based on recognition and is expected to provide a usable security for PDAs by adopting a similar drawing inputs method in recall-based systems.

## III. OUR SCHEME

The proposed shoulder-surfing resistant scheme can be considered as an improvement of Story, as it keeps most of the advantage of Story and achieves stronger security. Like Story, our scheme is based on recognition, an easier memory task than recall [21, 22], and suggests users to create a story for sequence retrieval. Instead of direct input, it depends on users drawing a curve across their password images (pass-images) in order. The curve containing both pass-images and decoys guards against shoulder-surfing attacks by human observation. Using a drawing input method, our scheme is designed to empower users to log in their mobile devices quickly. The following paragraphs describe the design and present a prototype.

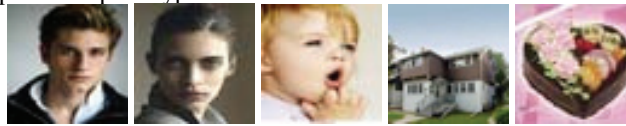

Figure 1.  The original five pass images.

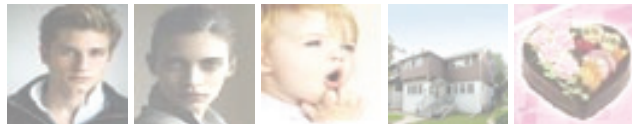

Figure 2.  The corresponding degraded images.

Our scheme uses a set of images gathered from http://FordModels.com and http://images.google.com. The images covering objects, places and people are carefully selected to motivate users' imagination and resized to have identical aspect ratios (Figure 1). To create a password, a user orderly chooses several images from the set as his/her pass-images (Figure 3). The user can remember the connection between the pass-images by mentally constructing a story. For confirmation, the user should draw a curve to cross his/her pass-images in right order without lifting the stylus from the drawing surface. With the similar interface, the confirmation phase also gives the user a chance to familiarize the authentication process.

During the authentication, a degraded version of the images is randomly arranged on the screen. These degraded images are formed by increasing brightness and reducing contrast of the original images (Figure 2). User must recognize his/her degraded pass-images and draw a curve to orderly cross them. The curve thus passes through both pass-images and other random images used to confuse the attacker. To avoid revealing pass-images (considering a user usually starts with the first pass-image and ends with the last), the drawing must begin with a given random image (head image) determined by the system and end likewise with another (tail image). Moreover, the majority of the drawing trace will be cleared away as the stylus being sliding and only the tail part will be kept to show the current location of the stylus. Without a visual clutter, the user can draw a complicated curve to bewilder peepers. At the same time, the trace length measured in the number of images crossed is limited within a tolerance in view of the random guess and brute force attacks. To ensure the legal user logging accurately, the tolerance must be greater than the sum of the minimum distance between sequential pass-images. The tolerance is adjustable and roughly defined by the maximum length of drawing trace $L$ in our prototype.

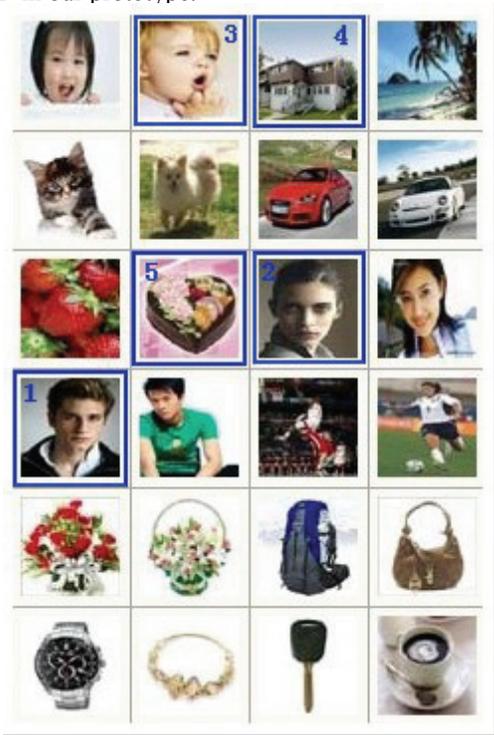

Figure 3. Graphical interface during the password creation.

Figure 3 shows a prototype of our scheme, which uses a template of 24 identically sized images, grouped into a $4 \times 6$ matrix. In Figure 3, the pass-images labeled by blue rectangles are in identical order with that in Figure 1. A hidden story may be "a couple and the son want to share a cake at their house". To log in, a user should draw a curve to orderly cross the pass-images from the given head image to the tail image. A possible drawing to successfully log in the scheme is shown in Figure 4 where the head is labeled by red rectangle and the tail green.

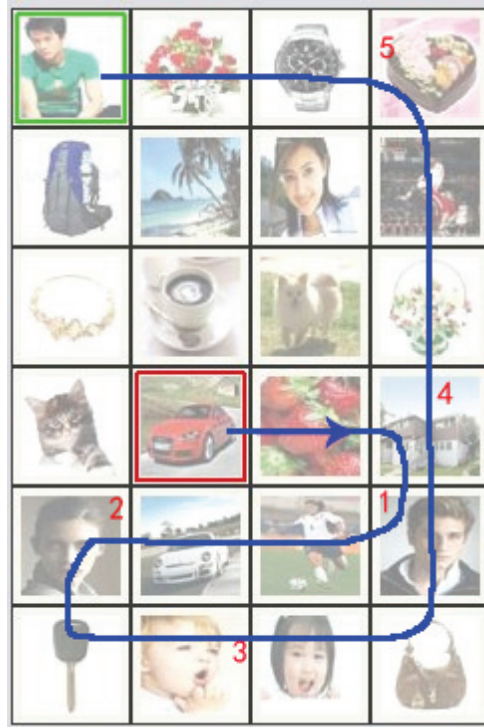

Figure 4. A possible drawing trace with length being 19.

The whole drawing trace is not actually visible and only a small tail part, about half the width of an image, will be displayed to indicate the current position. In default, the maximum length of drawing trace $L$ is defined in terms of the number of images in width $w$ and in length $l$ as well as the number of pass-images $n$ by: $L = (w+l) \cdot (n+1)$. Therefore, in Figure 4, $L = (4+6) \times (5+1) = 60$. Beyond its shoulder-surfing resistant properties, the prototype has a larger password space than that of commonly used four-digit alphanumeric PINs and thus more secure.

IV. USER STUDY

A. Methodology

To discuss the usability of CDS, we invited twenty university students (10 males and 10 females) to our lab. They were in the age range of 20 to 30 and none of them were familiar with graphical password schemes. We conducted a between-subjects design to benchmark the usability against that of Story [9]. So, half of the subjects were assigned to the CDS group (using the prototype of CDS) and half to the control group (using Story with the identical deployment in Figure 3). For both group, each participant was required to select five images as his/her pass-images.

Participants carried out the usability study individually in two sessions, an initial session (session 1) and a follow-up session (session 2) one week later.

In the first session, each participant was asked to read an instruction document. This provided information of their activities on the experiments and helped them know how to use Story or CDS. Then both groups orderly chose five images as pass-images and created a story to remember them. During the following confirmation, people in the CDS group checked their passwords and performed a drawing practice while the counterparts simply reentered the passwords. After a short delay (about 5 minutes), all the participants should log in repeatedly until ten successful logins were achieved.

One week later, at Session 2, all the participants returned to the lab and tried to log in with their previously created passwords within three attempts.

For each group, the corresponding scheme was instrumented to collect data on the number of correct and incorrect logins, the time for each login, and the total time for registration. We also recorded the length of drawing trace for CDS group when logging in. In addition, the story made in the creation was collected through a questionnaire.

*B. Results*

Two types of statistical tests were used to evaluate whether differences in the data occurred by conditions or by chance. We used t-tests (two tails) to compare the variance of the means between two groups and F-tests (one tail) to compare the mean login time within a single group. In the following, a value of $p<0.05$ indicates that the test revealed statistically significant difference and "not significant" (i.e. $p>0.05$) indicates that the differences were not statistically of significance.

*1) Password generation*

Table 1 shows the time to create a password in both groups. All the participants could create a password within 2 minutes correctly, wherein sixteen participants accomplished the registration in one minute and the other four participants took from 75s to 107s (seconds). We found that there was a great difference between the fastest participant and the slowest one (16s vs. 107s). It was reported that some people found it simple and quick to construct a story while some took a long to conceive, indicating that the time differences across all participants resulted probably from the inherent association ability. The mean time in CDS group was a bit longer than that in control group (49.5s vs. 42.9s), but not to a statistically significant level. As such, the additional familiarization with the drawing in CDS comes at the cost of longer time to carry out the registration.

TABLE I. PASSWORD GENERATION TIME (SECONDS) IN EACH GROUP.

| Group | Avg. | t-test | S.d. | Max | Min |
|---|---|---|---|---|---|
| CDS | 49.5 | Not significant | 26.7 | 107 | 19 |
| Story | 42.9 | | 24.3 | 93 | 16 |

*2) Login*

At session 1, all participants achieved the criterion of ten correct logins. In CDS group, seven of the participants accomplished the criterion in ten attempts with no errors. The other three participants made a total of four incorrect logins. The mean success login rate is 96.5% (Standard Deviation, StdDev=5.98). In control group, eight participants made no mistake in ten login attempts and each of the other made one incorrect login. The mean success rate is 98.2% (StdDev=3.83), higher than that of CDS group. However, it is useful to note that in practical terms, this was only a difference of two incorrect responses and should be treated with caution.

TABLE II. LOGIN TIME (SECONDS) FOR ALL CORRECT INPUTS AT SESSION 1.

| Group | Avg. | t-test | S.d. | Max | Min |
|---|---|---|---|---|---|
| CDS | 13.7 | P<0.01 | 5.97 | 45 | 7 |
| Story | 9.2 | | 4.26 | 32 | 4 |

The mean time for correct password inputs was also analyzed. As shown in Table 2, the participants in CDS group took more time to log in than their counterparts in control group on average (13.7s vs. 9.2s). A t-test yields a result of t=3.91, p<0.01(two tails), indicating that there was statistically significant difference between two conditions.

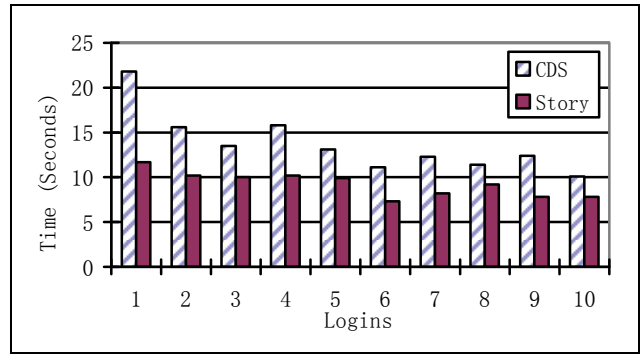

Figure 5. A possible drawing trace with length being 19.

Figure 5 shows the mean times for input of ten correct passwords in both groups. We can see that both groups kept a downward trend in time to input the password and the trend in CDS group looked more salient. Moreover, the difference between two groups became more and more slight. For CDS group, an F-test (one tail) yields a result of F=4.11, p<0.01, showing a significant decrease in time to input the password over the ten logins. For the control group, the result of F-test (one tail) indicated that there were no significant differences over the ten logins. It can be concluded that users will become adroit at CDS with frequent use.

TABLE III. LOGIN TIME (SECONDS) AT SESSION 2.

| Group | Avg. | t-test | S.d. | Max | Min |
|---|---|---|---|---|---|
| CDS | 19.8 | Not significant | 11.7 | 48 | 5 |
| Story | 23.1 | | 16.2 | 60 | 7 |

One week later, 19 participants were able to correctly recall their passwords and one participant could identify all of his pass-images but forgot the order. 80% of the CDS group and 60% of the control group could log in successfully with one attempt. The remainder of the CDS group succeeded in two attempts while one participant in the control group failed and the others succeeded within three attempts. It appears that the CDS group performed better than the control group in the long-term recall.

With the superiority in recall rate, the CDS group had an advantage over the control group in the mean time to log in (19.8s vs. 23.1s). Table 3 shows the results which were the total time to log in with three chances instead of a single correct attempt in Table 2. It should be noted that the control group (13.5s, StdDev=5.65) was still faster than the CDS group (15.25s, StdDev=5.31) when only evaluating mean time of the successful login in one attempt.

## V. Discussion

### A. Login time

It took 13.7 seconds to log in our scheme on average, an advantage over other shoulder-surfing resistant schemes [11, 19, 20]. Through the identical association mnemonics, CDS has a similar memorability in comparison to Story. However, CDS was not as good as Story in respect of login time. We took a closer look at this issue and found that the offset arose probably from two aspects.

On the one hand, the degrade images in the prototype of CDS seemed to impinge on the high efficiency to log in. As reported by some participants, it took more efforts for users to find out a degraded image with less color contrast. The random head and tail images used to protect the first and last pass-images also had a negative influence on the usability. For example, at the initial stage, participants were prone to forget starting with the head and ending with the tail, so they must redraw a curve. It was worthwhile to note that this phenomenon faded away as the number of login increased, consistent with the downward trend in Figure 3.

On the other hand, we used a PC to simulate the performance of our prototype in PDAs and thus participants had to use a mouse instead of a stylus for input. As known, mouse-clicking was much more convenient than dragging, especially for users unfamiliar with mouse operation. A quick dragging was more likely to go beyond the input area than a clicking and some participants in CDS group did occasionally draw outside, never occurring in the control group. Furthermore, things would be different when the PDAs with a stylus were usable. Unlike the mouse, a stylus was easier to operate and the drawing method appeared to be more compatible with people's writing habit. Then, it was unlikely to draw beyond the input area and should be more comfortable to log in CDS quickly. However, we should conduct another experiment to check this conclusion.

### B. User behavior

From the viewpoint of security, the risks of CDS come largely from insecure user behavior. First, some participants tended to move the mouse to the first pass-image instead of the random head image. So the first pass-image was easily to be cracked. Second, we found that most of the participants kept the mouse on a pass-image for a second before dragging it to another. Final, hotspot was still a serious security problem. Figure 4 shows the most popular images selected by 11, 9 and 8 participants respectively. Figure 5 shows the least popular images over all participants. Each was selected only by one participant. It appeared that images in Figure 4 were easy to spur an association while the images in Figure 5 are hard. Therefore, the image set was of great importance and should be redesigned with more care.

### C. Drawing trace and image distribution

The tolerance of the drawing trace and the location of the pass-images, affected both the usability and security. A proper tolerance was necessary to prevent efficiently against random attempt and ensure the legal user logging accurately. In our experiment, it appeared that the default tolerance (60) was friendly for participants to pass the authentication since none of them exceeded the range. The mean length of drawing trace in authentication was 23.7 (StdDev=5.98) and range was from 14 to 40, indicating that the given tolerance of the trace length could be narrowed to enhance the security without impairing the usability. Instead of the accuracy, the location of the pass-images was closely related with login time. It took shorter to find pass-images when they were close to each other, similar to the finding in literature [11]. But, with a small portion of the window, the drawing would cross more pass-images and fewer random images, resulting in the pass-images easily being cracked. In addition, the distribution of the pass-images in location restricted the minimum of the trace length. A centered distribution correlated with a shorter length while a scatter with a longer length, a general phenomenon in our experiment. This revealed that it was better to set a relative tolerance than an absolute maximum length of drawing trace.

### D. Other

We found the CDS group was more likely to use a story CDS during the one-week recall, which was not consistent with the finding by Davis, et al [9]. This may be caused by the drawing practice, our experiment requirements or by chance. A wide user study is necessary to illustrate it in depth. In addition, our scheme can be adjusted to adapt to different security demands and application situations. For example, the input can be implemented by drawing several strokes in order rather than a single curve.

## VI. Conclusion

In this paper, a new shoulder-surfing resistant scheme CDS was proposed. It adopts a visual login technique that matches the capabilities and limitations of most handheld devices and provides a simple and intuitive way for users to authenticate. As such, it is an example of "usable security". The main contribution is that it overcomes a drawback of recall-based systems by erasing the drawing trace and introduces the drawing method to a variant of Story to resist shoulder-surfing. Usability testing of the CDS scheme showed that users were able to enter their passwords accurately and to remember them over time. In the first ten successful login, it took participants about 13.7 seconds to log in CDS at average and there was a gentle downward trend in time to input the password. One week later, all the participants recalled their CDS passwords correctly and spent approximately 19.8 seconds recalling and logging in.

Further work includes redesigning the image set, finding a proper tolerance for the drawing trace length, reducing the login time as well as conducting another user study where users are required to log in PDAs equipped with a stylus instead of a simulation on PCs. We also plan to investigate the entropy issue of pass-images and to study in more depth the effect of the association mnemonics on graphical passwords.

ACKNOWLEDGMENT

The authors would like to thank the reviewers for their careful reading of this paper and for their helpful and constructive comments. Project 60903198 supported by National Natural Science Foundation of China.